\documentclass[twocolumn,showpacs,preprintnumbers]{revtex4}
\usepackage{amssymb}
\usepackage{amsmath}
\usepackage{dcolumn}
\usepackage{bm}
\usepackage{graphicx}

\begin{document}

\title{Cavity QED implementation of multi-qubit refined Deutsch-Jozsa
algorithm}
\author{Wan Li Yang$^{1},$ Chang Yong Chen$^{2},$ Zhen Yu Xu$^{1,3},$ and
Mang Feng$^{1}$}
\email{mangfeng@wipm.ac.cn}
\affiliation{$^{1}$State Key Laboratory of Magnetic Resonance and Atomic and Molecular
Physics, Wuhan Institute of Physics and Mathematics, Chinese Academy of
Sciences, Wuhan 430071, China }
\affiliation{$^{2}$Department of Physics and Information Engineering, Hunan Institute of
Humanities, Science and Technology, Loudi 417000, China}
\affiliation{$^{3}$Graduate School of the Chinese Academy of Sciences, Bejing 100049,
China}

\begin{abstract}
We theoretically study a realization of multi-qubit refined Deutsch-Jozsa
(DJ) algorithm using resonant interaction of many Rydberg atoms with a
single-mode microwave cavity, in which the multi-qubit controlled phase
gates could be accomplished efficiently. We show how to achieve a
multi-qubit refined DJ algorithm in high-fidelity even in the case of weak
cavity decay. We argue that the required operations in our scheme are almost
within the present experimental possibilities.
\end{abstract}

\maketitle

Quantum algorithms have displayed the unusual power in improving
computational speed over their classical counterparts, due to computational
parallelism or interference effects. Among the most frequently mentioned
algorithms \cite{Shor, Gro, Deu}, the Deutsch-Jozsa (DJ) algorithm is the
simplest, but demonstrates the power of quantum mechanics by distinguishing
the constant functions from the balanced functions using only one-step
logical computation regardless of the input size. Experimentally, the
original DJ algorithm \cite{Deu} and/or its modified version (i.e., refined
DJ algorithm) \cite{Coll}\ have been implemented in nuclear magnetic
resonance (NMR) system \cite{NMR1,NMR2}, quantum dot \cite{QD}, linear
optical system \cite{Opt}, and trapped ions \cite{Ion}. Additionally, there
have been some theoretical proposals for achieving DJ algorithm using
trapped electrons \cite{Ele}, polyatomic molecules \cite{Mol}, atomic
ensembles \cite{Ens}, Josephson charge qubits \cite{Jose} and cavity quantum
electrodynamics (QED) \cite{C1,C2,C3}.

In this paper, we will focus on an implementation of multi-qubit refined DJ
algorithm with the Rydberg atoms simultaneously passing through a
single-mode microwave cavity system. Relevant experiments have been carried
out for the resonant interaction of two or three atoms with cavity mode \cite%
{Re}, the entanglement between two atoms in a microwave cavity \cite{En},
and the two-qubit phase gate \cite{Gate}. Based on the available techniques,
our scheme could have following favorable characters: (i) By a smart
encoding, we may accomplish the multi-qubit gating by one step, which helps
to achieve a straightforward and fast implementation of the refined DJ
algorithm. This makes the necessary operations more efficient with respect
to the previous ideas \cite{C2,C3} using two-qubit conditional gates; (ii)
As the DJ algorithm realized in our scheme is the refined version \cite{Coll}%
, the auxiliary qubits are not necessary, which might greatly simplify the
experimental requirement particularly in the case of scalability; (iii)
Although the resonant interaction is unavoidably affected by the cavity
decay, the fast implementation could effectively suppress the detrimental
effect in the case of the weak cavity decay.
\begin{figure}[tbp]
\includegraphics[width=7cm]{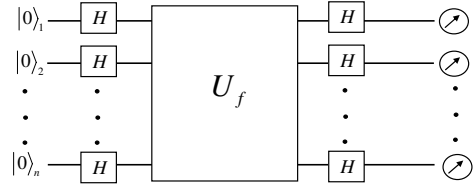}
\caption{Quantum circuit for the refined Deutsch-Jozsa algorithm, where $%
U_{f}$ is the $f$-controlled gate and the operation $H$ denotes the Hadamard
gate.}
\label{fig1}
\end{figure}

Let us first briefly review the main idea of the original DJ algorithm,
which can distinguish constant\ functions $f_{C}(x)$ from balanced\
functions $f_{B}(x)$ in an $N$-qubit system using only one query of
binary-valued function $f(x)$, namely, $f(x):\{0,1\}^{N}\rightarrow \{0,1\}$
\cite{Deu}. In the original DJ algorithm \cite{Deu}, the function is
characterized by the unitary operation $U\left\vert x\right\rangle
\left\vert y\right\rangle =$ $\left\vert x\right\rangle \left\vert y\oplus
f(x)\right\rangle ,$ where $x$ is an $N$-qubit input but the auxiliary qubit
$y$ must be prepared in the superposed state $(\left\vert 0\right\rangle
-\left\vert 1\right\rangle )/\sqrt{2},$ which results in the final
transformation
\begin{align}
U\left\vert x\right\rangle \left\vert y\right\rangle & =U\left\vert
x\right\rangle (\left\vert 0\right\rangle -\left\vert 1\right\rangle )/\sqrt{
2}  \notag \\
& \rightarrow (-1)^{f(x)}\left\vert x\right\rangle (\left\vert
0\right\rangle -\left\vert 1\right\rangle )/\sqrt{2}.
\end{align}
$\left\vert y\right\rangle $ seems superfluous because it keeps unchanged
during the operation process, although it actually plays a crucial\ role in
above DJ algorithm. This redundancy of the auxiliary qubit $y$ is fully
removed in the refined DJ algorithm \cite{Coll}, where the action of the $f$%
-controlled gate $\hat{U}_{f}^{(N)}$ can be denoted by $\hat{U}%
_{f}^{(N)}\left\vert x\right\rangle =$ $(-1)^{f(x)}\left\vert x\right\rangle
$ with $N$ being the qubit number. The basic quantum circuit to perform this
refined DJ algorithm is sketched in Fig. 1. For $N=2$, operation $\hat{U}%
_{f}^{(N)}$ can be specifically expressed as $\hat{U}_{f}^{(2)}=diag
\{(-1)^{f(00)},(-1)^{f(01)},(-1)^{f(10)},(-1)^{f(11)}\}$ in the state space
spanned by \{$\left\vert 00\right\rangle ,\left\vert 01\right\rangle
,\left\vert 10\right\rangle ,\left\vert 11\right\rangle $\}. It has been
shown that $\hat{U}_{f}^{(2)}$ can be reduced to a direct product of
single-qubit operations if $N\leq 2.$ This implies that the realization of
the DJ algorithm could be considerably simple. However, for $N\geq 3,$ the
situation will become much more intricate. In what follows, we will mainly
work on a three-qubit implementation of refined DJ algorithm. Our scheme is
directly extendable to many-qubit cases.

In the three-qubit case, along with two $f$-controlled gates $\hat{U}%
_{f_{C}}^{(3)}=\pm diag\{1,1,1,1,1,1,1,1\}$ corresponding to the constant\
functions $f_{C}(x),$ the number of $\hat{U}_{f_{B}}^{(3)}$ corresponding to
the balanced\ functions $f_{B}(x)$ is $C_{8}^{4}=70$. Actually, there are
only one $\hat{U}_{f_{C}}^{(3)}$ and $35$ nontrivial and distinct $\hat{U}
_{f_{B}}^{(3)}$ if we take the symmetry into account and neglect the overall
phase factors. So our task here is to implement 36 unitary transformations.

As the goal of the DJ algorithm is to differentiate the constant functions
from the balanced functions, instead of finding how $\hat{U}_{f}^{(N)}$
works specifically, we may simply consider the case below with one balanced
function and the corresponding $f$ -controlled operation%
\begin{equation}
\hat{U}_{f_{B1}}^{(3)}=diag\{1,-1,1,-1,-1,1,1,-1\},
\end{equation}%
where the state space is spanned by \{$\left\vert 000\right\rangle
,\left\vert 001\right\rangle ,\left\vert 010\right\rangle ,\left\vert
011\right\rangle ,\left\vert 100\right\rangle ,\left\vert 101\right\rangle
,\left\vert 110\right\rangle ,\left\vert 111\right\rangle $ \}.
\begin{figure}[tbph]
\centering\includegraphics[width=7cm]{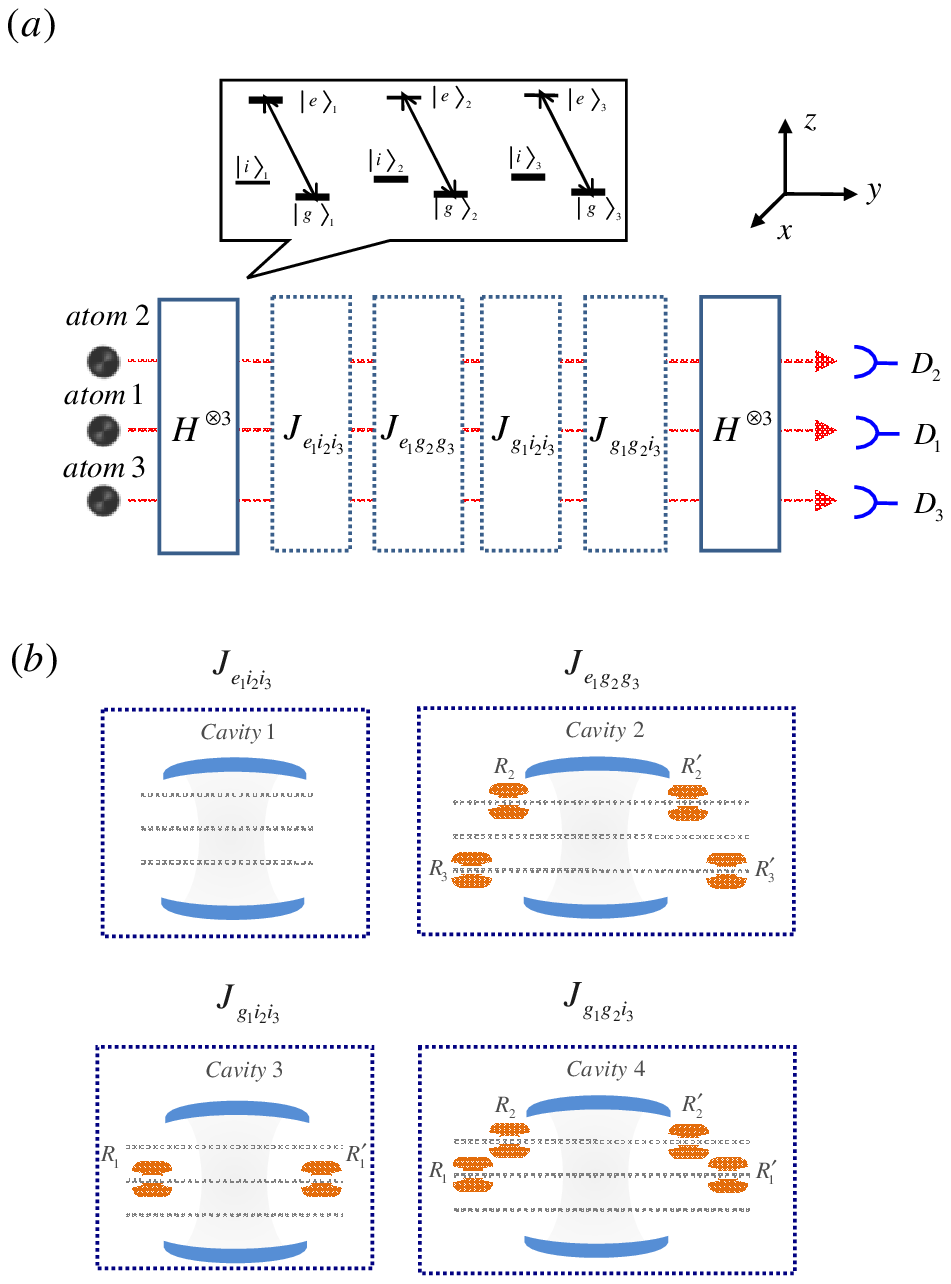}
\caption{(a) Schematic setup for implementing three-qubit refined DJ
algorithm in the case of the $f$-controlled gate is $\hat{U}%
_{f_{B1}}^{(3)}=diag\{1,-1,1,-1,-1,1,1,-1\}$, where the inset shows the
atomic level structure, and $D_{1}$, $D_{2}$, $D_{3}$ are state-selective
field-ionization detectors. (b) The detailed configuration of the unitary
operations $J_{e_{1}i_{2}i_{3}}$, $J_{e_{1}g_{2}g_{3}}$, $%
J_{g_{1}i_{2}i_{3}} $ and $J_{g_{1}g_{2}i_{3}}$. The cavity is a microwave
cavity sustaining a single mode with a standing-wave pattern along the $z$%
-axis. The atoms 1, 2 and 3 prepared in high-lying circular Rydberg states
are sent through the cavity with proper speed, resonantly interacting with
the cavity mode. $R_{i}(R_{i}^{^{\prime }})$ with $i=1,2,3$ denote Ramsey
zones for performing single-qubit rotations as explained in the text.}
\label{fig2}
\end{figure}

For clarity of following description, we first consider an ideal situation
without cavity decay, where three identical three-level atoms are input with
identical velocities and simultaneously interact with the single-mode vacuum
cavity field. The atomic internal states are denoted by $\left\vert
i_{j}\right\rangle ,$ $\left\vert g_{j}\right\rangle $ and $\left\vert
e_{j}\right\rangle ,$ with $\left\vert i_{j}\right\rangle $ decoupled from
other two states throughout our scheme due to large detuning, as shown in
Fig. 2 (a). In units of $\hbar =1,$ the effective Hamiltonian in the
interaction picture reads $H_{I}=\sum\nolimits_{j=1}^{3}\Omega
_{j}(a^{+}S_{j}^{-}+aS_{j}^{+}),$ where $\Omega _{j}$ is the coupling
constant of the $\allowbreak j$th atom to the cavity mode, $%
S_{j}^{+}=\left\vert e_{j}\right\rangle \left\langle g_{j}\right\vert $ and $%
S_{j}^{-}=\left\vert g_{j}\right\rangle \left\langle e_{j}\right\vert $\ are
the atomic spin operators for raising and lowering, respectively, and $a^{+}$
\ $(\allowbreak a)$\ is the creation (annihilation) operator for the cavity
mode.

The effective Hamiltonian $H_{I}$ is similar to the quantum computing model
of trapped ions in linear trap \cite{Yang}, where the multi-qubit controlled
phase flip $(CPF)$ gate $J_{\rho }=I-2\left\vert \rho \right\rangle
\left\langle \rho \right\vert $ with $I$ the identity matrix and $\left\vert
\rho \right\rangle =\left\vert 000\right\rangle ,\left\vert 001\right\rangle
,\cdot \cdot \cdot ,$ and $\left\vert 111\right\rangle ,$ could be achieved
with a high success probability and a high fidelity. In the present scheme,
we will try to move such an idea to cavity QED system for designing $CPF$
gate, which is essential to the $f$ -controlled operation $\hat{U}%
_{f}^{(N)}. $ Different from the time-varying coupling strength $\Omega
_{j}(t)$ in \cite{Yang}, however, the coupling strength in our scheme is
time-independent, i.e., $\Omega _{j}=\Omega _{0}\cos (2\pi z/\lambda
_{0})\exp (-r^{2}/w^{2})\sim \Omega _{0}\cos (2\pi z/\lambda _{0})$ \cite%
{Rev}, where $r$ is the distance of the atom away from the cavity center,
and $\lambda _{0} $ and $w$ are the wavelength and the waist of the cavity
mode, respectively. We use the same qubit definitions as in \cite{Yang},
that is, the logic state $\left\vert 0\right\rangle $ ($\left\vert
1\right\rangle $) of the qubit 1 is denoted by $\left\vert
g_{1}\right\rangle $ ($\left\vert e_{1}\right\rangle )$ of the atom 1; $%
\left\vert g_{2}\right\rangle $ and $\left\vert i_{2}\right\rangle $ \ of\
the atom 2 encode the logic state $\left\vert 0\right\rangle $ ($\left\vert
1\right\rangle $) of the qubit 2; the logic state $\left\vert 0\right\rangle
$ ($\left\vert 1\right\rangle $) of the qubit 3 is represented by $%
\left\vert g_{3}\right\rangle $ ($\left\vert i_{3}\right\rangle )$ of the
atom 3. If we assume the different atoms with coupling constants $\Omega
_{1}:\Omega _{2}:\Omega _{3}=1:10:10$ and the gating time $T_{0}=\pi /\Omega
_{1}$, we can obtain an approximate three-qubit $CPF$ gate $%
J_{111}=J_{e_{1}i_{2}i_{3}}=diag\{1,1,1,1,\alpha _{1},\alpha _{2},\alpha
_{3},-1\}$ in the computational subspace spanned by $\left\vert
g_{1}\right\rangle \left\vert g_{2}\right\rangle \left\vert
g_{3}\right\rangle ,\left\vert g_{1}\right\rangle \left\vert
g_{2}\right\rangle \left\vert i_{3}\right\rangle ,\left\vert
g_{1}\right\rangle \left\vert i_{2}\right\rangle \left\vert
g_{3}\right\rangle ,\left\vert g_{1}\right\rangle \left\vert
i_{2}\right\rangle \left\vert i_{3}\right\rangle ,$ $\left\vert
e_{1}\right\rangle \left\vert g_{2}\right\rangle \left\vert
g_{3}\right\rangle ,\left\vert e_{1}\right\rangle \left\vert
g_{2}\right\rangle \left\vert i_{3}\right\rangle ,\left\vert
e_{1}\right\rangle \left\vert i_{2}\right\rangle \left\vert
g_{3}\right\rangle ,\left\vert e_{1}\right\rangle \left\vert
i_{2}\right\rangle \left\vert i_{3}\right\rangle ,$ where $\alpha
_{i}=[\Omega _{1}^{2}\cos (\Theta _{i})+\Theta _{i}^{2}-\Omega
_{1}^{2}]/\Theta _{i}^{2}\approx 1$ with $i=1,2,3,$ and $\Theta _{1}=\sqrt{%
\Omega _{1}^{2}+\Omega _{2}^{2}+\Omega _{3}^{2}},$ $\Theta _{2}=\sqrt{\Omega
_{1}^{2}+\Omega _{2}^{2}},$\ $\Theta _{3}=\sqrt{\Omega _{1}^{2}+\Omega
_{3}^{2}}.$

Based on the gate $J_{e_{1}i_{2}i_{3}},$ we construct the $f$-controlled
gate $\hat{U}_{f_{B1}}^{(3)}$ by a straightforward way, i.e.,
\begin{equation}
\hat{U}
_{f_{B1}}^{(3)}=J_{e_{1}i_{2}i_{3}}J_{e_{1}g_{2}g_{3}}J_{g_{1}i_{2}i_{3}}J_{g_{1}g_{2}i_{3}},
\end{equation}
where$\ J_{e_{1}g_{2}g_{3}}=J_{100}=\sigma _{x,3}\sigma
_{x,2}J_{e_{1}i_{2}i_{3}}\sigma _{x,2}\sigma _{x,3},$ $%
J_{g_{1}i_{2}i_{3}}=J_{011}=S_{x,1}J_{e_{1}i_{2}i_{3}}S_{x,1}$ and $%
J_{g_{1}g_{2}i_{3}}=J_{001}=\sigma
_{x,2}S_{x,1}J_{e_{1}i_{2}i_{3}}S_{x,1}\sigma _{x,2},$ with $\sigma
_{x,j}=|i_{j}\rangle \langle g_{j}|+|g_{j}\rangle \langle i_{j}|$ with $%
j\neq 1,$ and $S_{x,1}=|e_{1}\rangle \langle g_{1}|+|g_{1}\rangle \langle
e_{1}|.$ The other four indispensable $CPF$ gates could also be created as $%
J_{e_{1}i_{2}g_{3}}=J_{110}=\sigma _{x,3}J_{e_{1}i_{2}i_{3}}\sigma _{x,3},$ $%
J_{e_{1}g_{2}i_{3}}=J_{101}=\sigma _{x,2}J_{e_{1}i_{2}i_{3}}\sigma _{x,2},$ $%
J_{g_{1}i_{2}g_{3}}=J_{010}=\sigma
_{x,3}S_{x,1}J_{e_{1}i_{2}i_{3}}S_{x,1}\sigma _{x,3},$ $%
J_{g_{1}g_{2}g_{3}}=J_{000}=\sigma _{x,3}\sigma
_{x,2}S_{x,1}J_{e_{1}i_{2}i_{3}}S_{x,1}\sigma _{x,2}\sigma _{x,3}.$ Using
these eight $CPF$ gates, we could construct other $35$ $f$-controlled gates $%
\hat{U}_{f_{Bm}}^{(3)}$ $(m=1,2,\cdot \cdot \cdot ,35),$ with each $\hat{U}%
_{f_{Bm}}^{(3)}$ involving four different $CPF$ gates.

Along with the above-mentioned $CPF$ gates, three-qubit Hadamard gates $%
H^{\otimes 3}=\prod\nolimits_{i=1}^{3}H_{i}$ should be performed to encode
the input and decode the output, respectively, as depicted in Fig. 2(a),
where $H_{i}=[(\left\vert 0\right\rangle _{i}+\left\vert 1\right\rangle
_{i})\left\langle 0\right\vert _{i}+(\left\vert 0\right\rangle
_{i}-\left\vert 1\right\rangle _{i})\left\langle 1\right\vert _{i}]$ $/\sqrt{%
2}$ is the Hadamard gate acting on the $\allowbreak i$th atom. These gates
could be achieved using external microwave pulses. As a result, a full
three-qubit refined DJ algorithm is available. Taking the $f$-controlled
gate $\hat{U}_{f_{B1}}^{(3)}$ as an example, we have designed a three-qubit
refined DJ algorithm setup in Fig. 2, where three Rydberg atoms prepared in
the state $\left\vert \Psi _{0}\right\rangle =\left\vert
g_{1}g_{2}g_{3}\right\rangle $ are initially encoded by the three-qubit
Hadamard gate $H^{\otimes 3}$ \cite{explain}$,$ and then\ sent through the
cavity with the identical velocities. The implementation could be simply
described as,
\begin{equation*}
\left\vert \Psi _{0}\right\rangle \;\underrightarrow{H^{\otimes 3}\hat{U}
_{f_{Bm}}^{(3)}H^{\otimes 3}}
\end{equation*}
\begin{align}
& \left\vert \Psi _{fm}\right\rangle =\ell \{A_{m}\left\vert
001\right\rangle +B_{m}\left\vert 010\right\rangle +\allowbreak
C_{m}\left\vert 011\right\rangle +D_{m}\left\vert 100\right\rangle  \notag \\
& +E_{m}|101\rangle +F_{m}\left\vert 110\right\rangle +G_{m}\left\vert
111\right\rangle \};
\end{align}
\begin{equation}
\left\vert \Psi _{0}\right\rangle \;\underrightarrow{H^{\otimes 3}\hat{U}
_{f_{C}}^{(3)}H^{\otimes 3}}\;\left\vert \Psi _{f}\right\rangle =\pm
\left\vert 000\right\rangle ,
\end{equation}
where $\ell $ is a normalized coefficient with $A_{m},B_{m},\cdot \cdot
\cdot ,$ and $G_{m}$ being $0$ or $\pm 1$. It implies that, if $f(x)$ is
constant, the state of the atoms becomes $\left\vert 000\right\rangle ;$ But
if $f(x)$ is balanced, the state of the atoms becomes a superposition state $%
\left\vert \Psi _{f}\right\rangle _{m}$, excluding the component $\left\vert
000\right\rangle .$ So we can efficiently determine whether the function is
constant or balanced by a collective measurement on the output state of the
three atoms.

Under the assumption of weak cavity decay that no photon actually leaks out
of the microwave cavity during our implementation, we may reconsider our
scheme using the quantum trajectory method \cite{Rev},
\begin{equation}
H_{D}=\sum\nolimits_{j=1}^{3}\Omega _{j}(a^{+}S_{j}^{-}+aS_{j}^{+})-i\frac{
\kappa }{2}a^{+}a,
\end{equation}
where $\kappa $ is the cavity decay rate. When we choose the atom-cavity
interaction time $\allowbreak T_{D}=\pi /A_{1\kappa }$ with $A_{1\kappa }=
\sqrt{\Omega _{1}^{2}-\kappa ^{2}/16}$ and meet the condition $\Omega
_{1}:\Omega _{2}:\Omega _{3}=1:10:10,$ the approximate three-qubit $CPF$
gate\ in the decay case becomes $\allowbreak $ $J_{e_{1}i_{2}i_{3}}^{^{
\prime }}=diag\{1,1,1,1,c_{1},c_{2},c_{3},c_{4}\}$ \cite{Exp}, and the
gating time becomes $T_{D}=\pi /\sqrt{\Omega _{1}^{2}-\kappa ^{2}/16}$.

For an initial state $\left\vert \Psi _{0}\right\rangle =\left\vert
000\right\rangle $, after the operations in Fig. 2, the output state of the
three atoms in the decay case is given by $\left\vert \Psi _{fm}^{^{\prime
}}\right\rangle =R_{m}\left\vert 000\right\rangle +A_{m}^{^{\prime
}}\left\vert 001\right\rangle +B_{m}^{^{\prime }}\left\vert 010\right\rangle
+\allowbreak C_{m}^{^{\prime }}\left\vert 011\right\rangle +D_{m}^{^{\prime
}}\left\vert 100\right\rangle +E_{m}^{^{\prime }}|101\rangle
+F_{m}^{^{\prime }}\left\vert 110\right\rangle +G_{m}^{^{\prime }}\left\vert
111\right\rangle $, where $R_{m}\approx 0$ and $A_{m}^{^{\prime
}},B_{m}^{^{\prime }},\cdot \cdot \cdot ,G_{m}^{^{\prime }}$ are slightly
deviated from $A_{m},B_{m},\cdot \cdot \cdot ,G_{m}.$ In Fig. 3(a), we
demonstrate the fidelity according to following relation $F_{m}=\overline{%
\langle \Psi _{fm}|\Psi _{fm}^{^{\prime }}\rangle \langle \Psi
_{fm}^{^{\prime }}|\Psi _{fm}\rangle }$ \cite{Fe}, in which we have also
considered the influence from the deviation $\Delta $ due to slightly
different atomic velocities.

In addition to the imperfection considered above, there are other noise
effects need to investigate, such as the resonant dipole interaction between
two neighboring Rydberg atoms. Since the typical dipole moment of Rydberg
atoms is about several hundred $q_{e}a_{0},$ where $a_{0}$ is the Bohr
radius and $q_{e}$ is electron charge, we can make a calculation of the
dipole coupling strength $\delta $ between two neighboring Rydberg atoms
being in the same range as the cavity decay rate $\sim KHz.$ Considering the
system Hamiltonian (Eq. (6)), we should also take an additional term $%
\sum\nolimits_{j=1}^{N}\delta |e_{j}e_{j+1}\rangle \langle e_{j}e_{j+1}|$
into account to assess the influence of the nearest-neighbor dipole-dipole
interaction.

In the present of cavity decay, we have plotted in Fig. 4 the fidelity of $%
\hat{U}_{f_{B1}}^{(3)}$ in a three-qubit refined DJ algorithm versus the
parameter $\delta /\Omega _{1}$ and different cavity decay rates. As shown
in Fig. 4, as long as these dipole interactions are weak, our scheme can
still achieve a high fidelity. In a realistic experiment, the situation
would be more complicated than our consideration above. So to carry out our
scheme efficiently and with high fidelity, we have to suppress these
above-mentioned imperfect factors as much as we can.

\begin{figure}[tbph]
\centering\includegraphics[width=10cm]{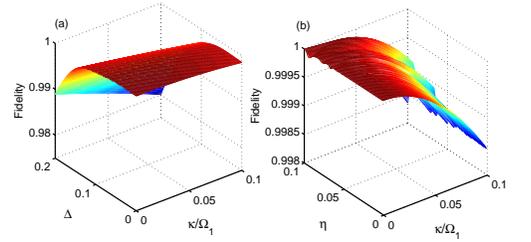}
\caption{(a) Fidelity of $\hat{U}_{f_{B1}}^{(3)}$ in a three-qubit refined
DJ algorithm versus the parameter $\protect\kappa /\Omega _{1}$ and the
deviation of gating time $\Delta $. (b) Fidelity of $\hat{U}_{f_{B}}^{(4)}$
in a four-qubit refined DJ algorithm versus the coupling ratio $\protect\eta %
=\Omega _{1}/\Omega $ and the parameter $\protect\kappa /\Omega _{1}$.}
\end{figure}

Extending to the many-qubit case, we could construct the $N$-qubit $CPF$
gate $\hat{V}_{\left\vert 111\cdot \cdot \cdot \cdot \cdot 1\right\rangle }=$
$diag\{1,1,1,\cdot \cdot \cdot ,-1\}$ by meeting the condition for coupling
constants $\Omega _{1}\ll \Omega _{2}=\Omega _{3}=\cdot \cdot \cdot =\Omega
_{N}=\Omega $ and by keeping the gating time $T_{D}=\pi /\sqrt{\Omega
_{1}^{2}-\kappa ^{2}/16}$ unchanged. This implies that the $CPF$ gate could
be carried out by a constant time irrelevant to the qubit number, which is
favorable for a scalable DJ algorithm in cavity QED system. Since the
single-qubit operation takes negligible time in comparison with that for the
many-qubit $CPF$ gate, we may roughly omit the single-qubit operation time,
which yields running time of the $N$-qubit refined DJ algorithm to be $%
T_{(N)}=$ $2^{N-1}\times T_{D}.$ Fig. $3(b)$ demonstrates an example of
four-qubit $f$-controlled gate $\hat{U}_{f_{B}}^{(4)}=diag%
\{1,-1,1,-1,-1,1,-1,1,1,-1,-1,1,1,-1,-1,1\},$ constructed by

\begin{eqnarray}
\hat{U}_{f_{B}}^{(4)}
&=&J_{g_{1}g_{2}g_{3}i_{4}}J_{g_{1}g_{2}i_{3}i_{4}}J_{g_{1}i_{2}g_{3}g_{4}}J_{g_{1}i_{2}i_{3}g_{4}}
\notag \\
&&J_{e_{1}g_{2}g_{3}i_{4}}J_{e_{1}g_{2}i_{3}g_{4}}J_{e_{1}i_{2}g_{3}i_{4}}J_{e_{1}i_{2}i_{3}g_{4}}.
\end{eqnarray}

\begin{figure}[tbph]
\centering\includegraphics[width=4.8cm]{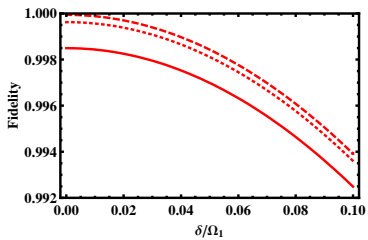}
\caption{Fidelity of $\hat{U}_{f_{B1}}^{(3)}$ in a three-qubit refined DJ
algorithm versus the parameter $\protect\delta /\Omega _{1}$ and the cavity
decay rate $\protect\kappa $, where the solid, dotted, and dashed curves
represent the cases of $\protect\kappa =\Omega _{1}/10,$ $\protect\kappa %
=\Omega _{1}/20,$ and $\protect\kappa =\Omega _{1}/50$, respectively.}
\label{fig4}
\end{figure}

We briefly address the experimental feasibility of our scheme by considering
three high-lying Rydberg atomic levels with principal quantum numbers 49, 50
and 51 to be levels $\left\vert g\right\rangle $, $\left\vert i\right\rangle
$ and $\left\vert e\right\rangle ,$ respectively. Based on the experimental
numbers reported in \cite{Gle}, the coupling strength at the cavity centre
could be $\Omega _{0}=2\pi \times 51$ KHz, and the Rydberg atomic lifetime
is $30$ ms. Specifically, assuming $\Omega =2\pi \times 51$ KHz, $\Omega
_{1}=2\pi \times 5.1$ KHz, and $\kappa =10^{-3}\Omega _{1},$ direct
calculation shows that the time for a single $N$-qubit $CPF$ gate is $%
T_{D}=\pi /\sqrt{\Omega _{1}^{2}-\kappa ^{2}/16}\approx 98$ $\mu $s, which
is much shorter than either the cavity decay time, i.e., $2\pi /\kappa
\approx 0.2$ s, or the Rydberg atomic lifetime. Actually, the lifetime of
the photon in the superconducting cavity has reached $0.5$ s recently \cite%
{Gle}. To make the quantum trajectory menthod workable, we require our
implementation time to be much shorter than the cavity decay time, which
yields the condition $\kappa \leqslant \Omega _{1}/2^{N-2}$. Considering the
values listed above, we find that N could be 9.

In current microwave cavity experiments \cite{Rev}, the intra-atom
interaction occurs in the central region of the cavity with the Rabi
frequency $\Omega _{j}=\Omega _{0}\cos (2\pi z/\lambda _{0})$ \cite{Rev}. To
meet the condition of $\Omega _{1}=\Omega /10$, the $N$ atoms should be sent
through the cavity with the first atom going along the $\allowbreak y$-axis
deviating from the nodes by $(\lambda _{0}/2\pi )\arccos (1/10)$, but other
atoms through the antinodes. Specifically, for the $N$ input atoms (suppose $%
N$ being an even number for simplicity), we have the tracks of the atomic
movement as $z_{1}=-(N/2)\lambda _{0}+(\lambda _{0}/2\pi )\arccos (1/10)$, $%
z_{2}=-(N/2-1)\lambda _{0},$ ...., $z_{N}=(N/2)\lambda _{0}$. This is of
course highly challenging with current experimental technology because we
have not yet found any experimental report for $N\geqslant 3$ atoms
simultaneously going through a microwave cavity. However, the two-atom
entanglement in a microwave cavity has been achieved using van der Waals
collision between the atoms \cite{En} under a non-resonant condition.
Moreover, our scheme could also be straightforwardly applied to other
quantum information processing candidate systems, e.g., the ion-trap-cavity
combinatory setup \cite{Trap} or cavity-embedded optical lattices confining
atoms \cite{Latt}, in which the atoms are localized very well.

Alternatively, the superconducting circuit QED \cite{Cir,Wal,Fink} would
become a more suitable candidate for implementing our scheme. In the circuit
QED \cite{Cir,Wal}, a number of superconducting qubits play the role of
artificial atoms and quantum bus is provided by a 1D superconducting
transmission line resonator. So the dynamics of a two-level qubit coupled to
a single mode of an electromagnetic field is also described by the
Jaynes-Cummings Hamiltonian, which is an essential requirement for the
present scheme. Implementing multi-qubit refined DJ algorithm in the circuit
QED has several merits that are worth mentioning here. First, the
experiments performed in the circuit QED set-up have demonstrated that long
coherence time and very strong coupling could be realized \cite{Wal}, which
could greatly reduce the gating time and noise effects in our scheme.
Second, the widely separated superconducting qubits are well localized in
the circuit QED, so full in-situ control over the qubit parameters and
system qubit Hamiltonian is easily achieved. Third, the strong coupling
between the field in the resonator and the qubits can be used to perform a
single-shot high efficiency quantum nondemolition (QND) readout of the state
of the qubits without the need for additional signal ports.

In conclusion, we have proposed a potentially practical scheme for realizing
a multi-qubit refined DJ algorithm by resonant interaction of Rydberg atoms
in a microwave cavity. We have estimated the influence from the cavity decay
on our scheme and shown that our scheme could be achieved efficiently to
distinguish the balanced functions from the constant functions with high
fidelity. We argue that our present scheme would be helpful for
demonstration of refined DJ algorithm at large scale using cavity QED
devices.

This work is supported by NNSF of China under No. 10774163 and No. 10774042.


\begin{thebibliography}{99}
\bibitem{explain} For simplicity, we assume here that we could individually
address the atoms for implementing Hadamard gates. In real experiments,
however, due to the long wavelength of the microwave, we have to employ
inhomogeneous electric fields for single-qubit operations \cite{Yama}.
\end{thebibliography}
\end{document}